# STRATEGY PATTERN: PAYMENT PATTERN FOR INTERNET BANKING

[1]A. Meiappane, [2]J. Prabavadhi and [3]V. Prasanna Venkatesan
[1]Research Scholar, [3]Associate Professor, Pondicherry University, Puducherry
[2]PG Student, Department of CSE, Sri Manakula Vinayagar Engineering College, Puducherry
E-mails: [1]auromei@yahoo.com, [2]it_praba@yahoo.co.in, [3]prasanna_v@yahoo.com

*Abstract:* This paper brings out the design patterns according to the various services involved in internet banking. The Pattern oriented Software Architecture uses the pattern which eliminates the difficulty of reusability in a particular context. The patterns are to be designed using BPM (Business Process Model) for effective cross cutting on process level. For implementing the above said BPM, the Internet banking has been taken to implement the pattern into it. The Analysis and identification of various processes in Internet Banking have been done, to identify the effective cross cutting features. With this process the pattern has been designed, as a reusability component to be used by the Software Architect. The pattern help us to resolve recurring problems constructively and based on proven solutions and also support us in understanding the architecture of a given software system. Once the model is finalized by analyzing, we found payment in the process of internet banking has a strategy pattern.

## 1. INTRODUCTION

In software engineering, a design pattern is a general reusable solution to a commonly occurring problem within a given context in software design. A design pattern is not a finished design that can be transformed directly into code. It is a description or template for how to solve a problem that can be used in many different situations and it is also called as a blue-print of how to solve a problem. It is used to determine implementation faster and make code more readable to other programmers the design pattern is divided into three types: creational, structural, and behavioral. *Creational patterns* create objects for you rather than having you instantiate objects directly. This gives your program more flexibility in deciding which objects need to be created for a given case. *Structural patterns* help you compose groups of objects into larger structures, such as complex user interfaces or accounting data. *Behavioral patterns* help you define the communication between objects in your system and how the flow is controlled in a complex program. [14]

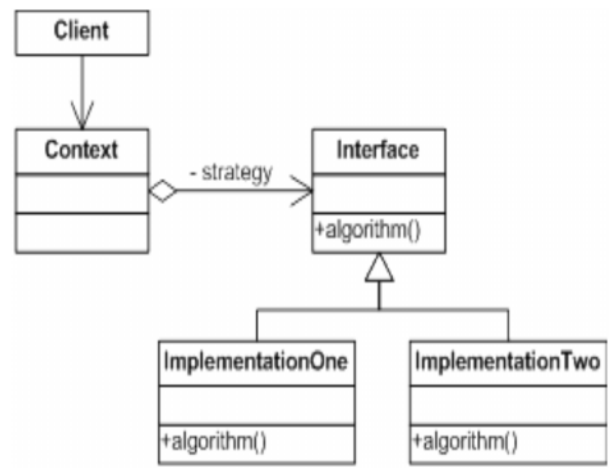

**Figure 1:** Structure of Strategy Pattern

The Strategy pattern has been concentrated here which is a behavioral pattern. The strategy pattern is defining a family of algorithms, encapsulate each one, and make them interchangeable. Strategy lets the algorithm vary independently from the clients that use it.

### 1.1. Software Architecture

Software architecture is the fundamental organization of a system, embodied in its components, their relationships to each other and to the environment, and the principles guiding its design and evolution.

### 1.2. Patterns

A particular recurring design problem that arises in specific design contexts, and presents a well-proven generic scheme for its solution. The solution scheme is specified by describing its constituent components, their responsibilities and relationships, and the ways in which they collaborate. Each pattern is a three-part rule, which expresses a relation between a certain context, a problem and a solution. Patterns can be applied to many different areas of human endeavor, including software development. [14]

### 1.3. Architectural Pattern

It is Standard design and it Express a fundamental structural organization for software system and provide set of predefined sub system and includes rules and guidelines for organizing the relationships between them.

### 1.4. Usage of patterns

- Comprehend existing systems; customize systems to fit user needs; and construct new systems.
- Identify and specify higher level abstractions
- Reinforce an architectural view of the system
- Explicitly address non-functional properties
- High availability and minimization of business risk
- They help in the construction of complex and heterogeneous software architectures
- They help to manage complexity

### 2. LITERATURE SURVEY

Opportunistic Linking is a pattern used in shopping websites quite similar to Advising, when applying this pattern it is advising the customer. However, the intent is different. While opportunistic linking tries to keep the user inside the store by giving him new ideas to buy, Advising helps him choose what he wants. [1]

The crosscutting concern problem causes the code related to a concern to be scattered across the program, and often tangled with the code related to other concerns. Several studies indicate that modularizing crosscutting concerns improves software quality, providing indirect proof that crosscutting hurts modularity. Unfortunately, there is little guidance for finding crosscutting concerns, and determining when it is profitable to modularize them. Before we can go about reducing crosscutting code to improve modularity, we must first determine *what* the concerns of the program are (*concern identification*) and *where* they manifest in the program text. Only then can we classify the concerns as crosscutting or no crosscutting. [2]

Two patterns for the infrastructure needed to build a web shop are the Catalog pattern and the Shopping Process pattern. The Catalog pattern organizes information about products in an efficient and flexible way. It also provides mechanisms to help customers make decisions. The goal of the Shopping Process pattern is to simplify the shopping process and to improve the efficiency and convenience of the buying Process. [4]

The development of the component based techniques has reached a certain maturity. The objective is to replace the traditional steps of development by two types of complementary processes: "design for reuse" and "design by reuse". In this context, this work lies within the scope of the "design by reuse" process. The encountered problem during the reuse is to find the best components which will be able to respond to the arising problem. Certainly, there are tools allowing a simple administration of the components but they showed certain limits since they did not take into account the user profile in the retrieval process. [5]

BPM is as a case-based workflow solution to support business processes, it can expose underlying complex business processes as well-defined interface and consume other business processes, service providers. BPM can support human to system, system to system, system to human or human to human interactions. [6]

As the electronic commerce (Be) activities become more and more diverse, it is very critical to provide the right information to the right customers. In EC environment, how to find the association rules among purchased items is very important. If this kind of information is provided to the web site manager, the performance of cross-selling should be improved. However, the web site managers may also pay attention to the navigation behaviors of customers. Association rule mining just purely considered the purchasing behaviors of customers and the managers' requirements. Therefore, it is very important to simultaneously consider the navigation behaviors and purchasing behaviors of customers, that is, combine the web traversal patterns and association rules to provide more information for the web site managers. The combinations of web traversal patterns and association rules are called web transaction patterns. In the following, we describe the definitions about web traversal patterns and web transaction patterns. [7]

Mining patterns are used in different aspects of Web applications such as their navigation topology, their interfaces. Some patterns are specific to particular application fields like e-commerce. In all cases, the web patterns are more similar to Alexandrian patterns than to patterns as they are not described as collaboration among objects but as navigable design structures. The context in which we mined these patterns is the Object-Oriented Hypermedia Design Method (OOHDM) approach [Schwabe98] though they can be obviously used with other methods. In OOHDM a Web application is conceived as a hypermedia view on an object model. This view comprises nodes (the objects that the user will navigate), links that conform the navigation topology, navigation contexts that represent sets to be navigated sequentially and different kinds of indexes. As with object-oriented patterns, hypermedia (and Web) patterns go beyond the naive use of the basic theoretical concepts. These patterns indicate how to build usable hypermedia topologies by creating elaborated structures. [8]

Survey using questionnaire about the customer utilization of services, internet websites: www.rbi.org.in,banknetindia.com. The survey data provided the needed input for the classification of the services required by users of different age group and different occupation groups. [9]

Patterns are reusable designs which can be used in other similar applications. Online share trading mainly deals with selling and buying of shares electronically. Online share trading is categorized into two types. One deals with the equity trading and the other deals with the options trading. Our focus is on equity trading. A pattern language is a set of related patterns in a specific domain. These patterns are useful to the developers while analyzing such systems. A software design pattern identifies a recurring problem and a solution describing them in a particular context to help developers understand how to create an appropriate solution [10].

Elements in the source are related to elements in the target. We use the term pattern as in design patterns, in the sense of being a general description of frequently encountered situations. Phrases as "one thing *with* respect *to* another thing". Some examples of source and target elements in the crosscutting pattern are the following: concern x module, concern x requirement, concern x architectural element, requirement x module, concern x implementation element. There is a mapping between source elements and target elements. The terms crosscutting, tangling and scattering are defined as special cases of these mappings. On one hand there is a relation or mapping between source elements and target elements. The mapping has a multiplicity. It could be 1:1 or 1: many. In case of 1: many mappings we have scattering, defined as follows: Scattering occurs when, in a mapping between source and target, a source element is related to multiple target elements. [11]

The good security and trust will ultimately increase the use of electronic commerce. In this paper, examine issues related to e-payment security from the viewpoint of customers. This study proposes a conceptual model that delineates the determinants of consumers' perceived security and perceived trust, as well as the effects of perceived security and perceived trust on the use of e payment systems. To test the model, structural equation modeling is employed to analyze data collected from 219 respondents in Korea. This research provides a theoretical foundation for academics and also practical guidelines for service providers in dealing with the security aspects of e-payment systems. [12]

Multi-factor authentication involves the use of more than one mode in authentication processes and is typically employed to increase security compared to a fixed password (knowledge-based mode). This research compared three different e Banking authentication processes, a two-layer password (1-factor) method and two alternative 2-factor solutions. The 2-factor processes used One-Time-Pass codes (OTPs) delivered either via a small, single-use device or by text message to a mobile phone. The three authentication methods were compared in a repeated-measures experiment with 141 participants. Three user groups were balanced in the experiment to investigate the effect of experience (current users of the service) on perceptions of usability. Majority of the participant sample perceived the 1-factor method they had most experience with as being the most secure and most convenient option. [13]

## 3. PATTERN MINING

Pattern mining is same as the data mining. Mining the pattern in internet banking by using the BPM (See Fig. 2). It is used to representing process involved in internet banking and is used to analyze the process then apply the crosscutting concerns approach in this model. The crosscutting concerns are the Aspect of program which affects the other concern. These concerns often cannot be cleanly decomposed from the rest of the system in both design and implementation and can the result in either scattering, tangling. The crosscutting concern is used to improve the modularity of the system.

In internet bank has lots of services and we analysis this services and find the crosscutting in the process level. The services are the account, Third party transfer, bill payment, credit card, debit card, Mutual fund.

Use of the crosscutting concern we found the payment pattern. The different types of payment available in the internet banking, we analysis the services and find the crosscutting in payment .The different types of payment is utility bill payment and credit card bill payment. We put these services under the payment pattern and this pattern is like the strategy design pattern. The pattern is used to increase the reusability and reduce the number of code and easily add new services to the internet bank with the small modification.

## 4. PATTERN CONSTRUCTION

A pattern for software architecture describes a particular recurring design problem that arises in specific design contexts and presents a well-proven generic scheme for its solution. The solution scheme is specified by describing its constituent components, their responsibilities and relationships, and the ways in which they collaborate."

Each pattern is a three-part rule, which expresses a relation between a certain context, a certain system of forces which occurs repeatedly in that context, and a certain software configuration which allows these forces to resolve themselves. Context section describes the situation in which the design problem arises. Problem section describes the problem that arises repeatedly in the context. And finally, solution section describes a proven solution to the problem.

Mining the payment pattern for internet banking. This pattern is like the strategy pattern. The pattern is constructed by using the elements are Name, Intent, Problem, and Solution (see Fig. 3).

The classes or object participating in this pattern are strategy is payment, it is used to declare an interface common to all supported algorithm. Client or customer uses this interface to call the algorithm defined by the concrete payment class. The concrete payment classes are credit card and utility bill payment. These classes implement the algorithm using the payment interface. Context is the client. it is configured with concrete payment object, maintains a reference to a strategy object and may define an interface that lets strategy access its data.

**Payment Pattern**

**Intent:** This pattern is used to pay money through online by choosing algorithm from family of algorithm at run time easily.

**Problem:** how the customer can easily choose the algorithm for money payment in internet banking.

**Solution:** for paying money through the internet banking, different types of payment method are used to pay money through internet banking such as credit card bill payment, Bill payment. Based on the customer needs they can choose any types of payment to pay bill or money through internet and they can be used interchangeably.

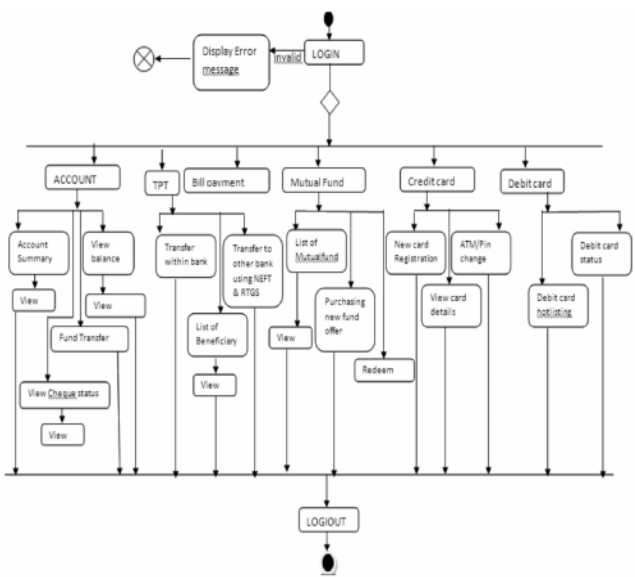

**Figure 2:** Business Process Model for Internet Banking

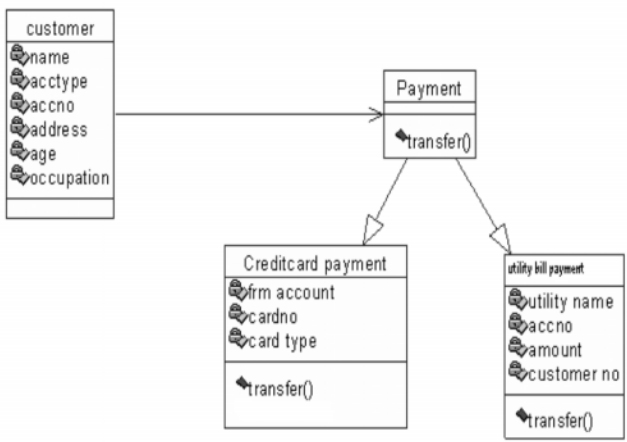

**Figure 3:** Strategy Pattern for Payment

## 5. CONCLUSIONS

The strategy pattern for internet banking has been presented over here. These patterns focus mainly on ways to solve usual problem. They provide hints to the Web application designer in order to make these applications more usable and effective both from the point of customers and designer of web application. These patterns help to improve the design and reusability. Further the research is proceeded on mining some more patterns for internet banking which will be a reusable component for internet Banking at a particular context.